\documentstyle[12pt]{article}
\normalbaselines
\begin{document}

\begin{center}
\vspace*{1.0cm}
{\LARGE {\bf Translocality and a Duality Principle in Generally Covariant
Quantum Field Theory\\}}

\vskip 1.5cm

{\large {\bf Hadi Salehi\footnote { e-mail address:
h-salehi@cc.sbu.ac.ir}}}

\vskip 0.5 cm

Department of Physics, Shahid Beheshti University , Evin Tehran 19839
Iran
\end{center}

\vspace{1cm}

\begin{abstract}
It is argued that the
formal rules of correspondence between local observation procedures
and observables do not exhaust the entire physical content of
generally covariant quantum field theory. This result
is obtained by expressing the distinguishing features
of the local kinematical
structure of quantum field theory in the generally covariant
context in terms of a translocal structure which
carries the totality of the nonlocal kinematical informations
in a local region.
This gives rise to a duality principle at the dynamical
level which emphasizes the significance of the underlying
translocal structure
for modelling a minimal algebra around a given point.
We discuss the emergence of
classical properties from this point of view.

\end{abstract}
\vspace{1cm}
\section{Introduction}
Quantum field theory studies
the properties of algebras which are
expected to give accurate mathematical descriptions of physical systems.
In  general, the manner in which one can extract informations
of direct physical relevance from the algebraic description is very subtle
because, for a given abstract algebra, there may exist in general
many (unitarily inequivalent)
representations in terms of operator algebras acting
on a Hilbert space.
Therefore, the basic problem of quantum field theory concerns the
characterization of physically admissible representations.\\
This problem is considerably simplified in the presence of space-time
symmetries. For example, in quantum field theory in Minkowski space,
because of the Lorentz symmetry, it is always possible to refer to a
representation containing the physical vacuum. A similar simplification could,
in principle, arise in any theory admitting at least a group of
space-time symmetry with a global time-like generator.\\
It turns out that in a generally covariant quantum field theory, because
of the dynamical role played by the space-time metric, no {\it a priori}
notion of space-time symmetry exists. Consequently, considerable
difficulties arise if one wants to characterize the physically admissible
representations.\\
Because of the lack of {\it a priori} space-time symmetries
in the generally covariant context, it is useful for the general treatment
of the basic problem of quantum field theory in that context to isolate
those features of the problem which can be discussed without reference
to any pre-assigned space-time symmetries.
It is perfectly possible that this may not resolve the problem
completely, nevertheless attempts
in this direction may provide important indications for understanding
the physical content of generally covariant quantum field theory.
The present paper contemplates a consideration of this issue within the scope
of the algebraic approach to quantum field theory [1].\\
We first briefly discuss the question of how general covariance can be
incorporated into the conventional framework of
quantum field theory [2]. The basic idea is to start with free
algebras, i.e. algebras which
are free from {\it a priori} relations.
The need for this is obvious, since otherwise
we have {\it a priori} no principle at hand ensuring that the
algebraic relations are
kept unchanged under the action of
an arbitrary space-time diffeomorphism. The general scheme we
shall now describe is a generalization of the scheme used in [2-5].\\
We consider a differentiable manifold ${\cal M}$
and assume the existence of a net of free algebras over ${\cal M}$ generated
by what we call kinematical procedures.
In specific terms we require
an intrinsic correspondence between each open set
$\cal O \in {\cal M}$ and a free involutive algebra ${\cal A}({\cal O})$ such that
the additivity
\begin{equation}
{\cal A}({\cal O})\subset {\cal A}({\cal O}^{\prime}),~~if~~
{\cal O}\subset {\cal O}^{\prime}
\label{1}  \end{equation}
holds.
The attribute 'intrinsic' means that
the principle of general
covariance is implemented by considering the group $Diff({\cal M})$
of all diffeomorphisms of the manifold
as acting by
automorphisms on the net of the algebras ${\cal A}({\cal O})$,
i.e. each diffeomorphism
$\chi\in Diff({\cal M})$ is represented by an automorphism $\alpha_{\chi}$ such that
\begin{equation}
\alpha_{\chi}({\cal A}({\cal O}))={\cal A}(\chi({\cal O}))
\label{1}  \end{equation}
holds.
Given such an intrinsic correspondence between open sets and algebras,
we call a self adjoint
element of ${\cal A}({\cal O})$ a kinematical procedure in $\cal O$. \\
We should emphasize that,
because there is no diffeomorphism invariant notion of
locality, it is by no means clear whether there is an {\it a priori}
correspondence between kinematical procedures
and local properties in the underlying
manifold. For example we may find a coordinate system in which
the kinematical procedures carry the global
properties of the entire manifold
in a "local domain", i.e., in a finite range of that coordinate system.
Typical examples of such coordinate systems in general relativity
are coordinate systems which compactify the structure of infinity.
Indeed, the exploration of the
question concerning the characterization of local
kinematical procedures is one of the basic tasks of the present analysis.
It will be dealt with in the next section.\\
There could be many kinematical procedures which are equivalent
with respect to the action of a physical system on them which is, in general,
expected to connect the kinematical procedures
with dynamical procedures (traditionally identified with observables).
Thus, the essential question
is how to identify the dynamical procedures of the net ${\cal O}\rightarrow
{\cal A}({\cal O})$ as suitable
equivalence classes of kinematical procedures?\\
For this aim, we first note that the precise
mathematical description of a physical system is given in terms of a
state which is taken to be
a positive linear functional over the
total algebra of kinematical procedures
${\cal A}:=\bigcup{\cal A}({\cal O})$. Given a state
$\omega$, one
gets via the GNS-construction a representation $\pi^{\omega}$ of $\cal A$
by an operator algebra acting on a Hilbert space ${\cal H}^{\omega}$ with
a cyclic vector $\Omega^{\omega}\in {\cal H}^{\omega}$. In the
representation $(\pi^{\omega}, {\cal H}^{\omega}, \Omega^{\omega})$
one can select a family of related
states on ${\cal A}$, namely those
represented by vectors and density matrices in ${\cal H}^{\omega}$. It
corresponds to the set of normal states of the
representation $\pi^{\omega}$, the so called folium of $\omega$.\\
Once a physical state $\omega$ has been specified, one can consider in each
subalgebra ${\cal A}({\cal O})$ the equivalence relation
\begin{equation}
A~\sim~B \longleftrightarrow
{\omega}^{\prime}(A-B)=0,~~ \forall  {\omega}^{\prime}\in{\cal F}^{\omega}.
\label{eq.}\end{equation}
Here ${\cal F}^{\omega}$ denotes the folium of the state $\omega$. The
set of such equivalence relations generates a two-sided ideal
${\cal I}^{\omega}({\cal O})$ in ${\cal A}({\cal O})$.
One can construct the
algebra of dynamical procedures ${\cal A}^{\omega}({\cal O})$ from
the algebra of kinematical procedures ${\cal A}({\cal O})$ by taking
the quotient
\begin{equation}
{\cal A}^{\omega}({\cal O})=
{\cal A}({\cal O})/{\cal I}^{\omega}({\cal O}).
\end{equation}
By this annihilation all the relevant relations between the dynamical
procedures can be characterized by the totality of elements in the
kernel of the representation $\pi^{\omega}$,
namely the total ideal ${\cal I}^{\omega}$
\begin{equation}
{\cal I}^{\omega}=\bigcup{\cal I}^{\omega}({\cal O}).
\end{equation}
This construction implies that the mapping from
kinematical procedures to dynamical procedures
becomes fundamentally state-dependent.
This aspect reflects one of the characteristic features of generally covariant
quantum field theory.\\
Crucial for further investigations is the realization
that a diffeomorphism
$\chi\in Diff({\cal M})$ can act as an automorphism $\alpha_{\chi}$ on the net
${\cal O}\rightarrow {\cal A}^{\omega}({\cal O})$ provided
\begin{equation}
\alpha_{\chi}({\cal I}^{\omega}({\cal O}))=
{\cal I}^{\omega}(\chi({\cal O}))
\label{2}\end{equation}
holds.
Any diffeomorphism satisfying this condition is called dynamical (or proper).
Otherwise it is called nondynamical (or improper). Nondynamical
diffeomorphisms can not be represented
as automorphisms on the algebra of the dynamical procedures.
For dynamical diffeomorphisms such a representation is possible. They
generate a group $G_{\omega}$ which is called in the following
the  dynamical group of $\omega$ and will be denoted by
$G_{\omega}$.
The elements of $G_{\omega}$ correspond to state-dependent
automorphisms of the algebra of dynamical
procedures with a pure geometric action. \\

\section{Local inertial sector}

One of the basic difficulty of the above scheme is that, in general,
the GNS-representation of a physical state can not unitarily be
fixed in an intrinsic manner, because
the structure of the total ideal ${\cal I}^{\omega}$ depends crucially
on the particular coordinates one uses. For example, starting
from the GNS-representation of a physical state one can obtain
another representation if the kinematical procedures are transformed by a
nondynamical diffeomorphism. The
representation obtained in this way may not be in the equivalence class
of the former because it may have a different kernel.
Thus a physical state will, in general,
provide us with a variety of unitarily inequivalent
representations depending on the nature of the coordinates that happened to have
been chosen for a given problem, and {\it a priori}
it is not known which representation is physical.\\
The problem can be addressed on various levels.
One possibility is to take a global point of view and
select the equivalence class of
representations for a physical
state $\omega$ as that for which the dynamical group $G_{\omega}$
is nontrivial and acts globally on the manifold $\cal M$.
The geometric action of this group would then determine
the nature of the equivalence class of coordinates
to which the representation refers.
These are coordinates which are related by the geometric action of the
dynamical group $G_{\omega}$. Such coordinates
may be considered as typical examples of global inertial coordinates. \\
A criterion of this type may be useful to analyze the particular
type of a physical theory resulting from the transition
from the generally covariant description of a physical state to
the special relativistic one. \\
For the description of a physical state in the generally covariant
context we shall formulate a local variant of the above criterion.
Specifically, we assume that, given a physical state $\omega$, we can assign
to any point $x\in \cal M$ a neighbourhood
${\cal O}^{\omega}_{x}$
so that by the restriction of the GNS- representation
$\pi^{\omega}$ to ${\cal O}^{\omega}_{x}$ a nontrivial
dynamical group $G_{\omega}$ is established which acts on
${\cal O}^{\omega}_{x}$. To emphasize the
individuality of the point $x$, we shall assume that
the geometric action of $G_{\omega}$ on ${\cal O}^{\omega}_{x}$
leaves the point $x$ invariant. In an alternative
formulation we shall require the invariance of the local ideal
${\cal I^{\omega}} ({\cal O}^{\omega}_{x})$ under the (nontrivial) action of
the dynamical
group $G_{\omega}$, namely
\begin{equation}
\alpha ({\cal I^{\omega}} ({\cal O}^{\omega}_{x}))=
{\cal I^{\omega}} ({\cal O}^{\omega}_{x}),
~~~ \forall \alpha\in G_{\omega}
\label{2eq1}\end{equation}
with $x$ being invariant  under to the geometric action of
$G_{\omega}$.
For any physical state $\omega$ this acts as a criterion to
select a characteristic local equivalence class of representations.
In symbols we
shall write for this local equivalence class
$\{\pi^{\omega}|{\cal O}^{\omega}_{x}\}$
and refer to it as a local inertial sector of a physical state $\omega$.
Correspondingly the equivalence class of
local coordinate systems to which $\{\pi^{\omega}|{\cal O}^{\omega}_{x}\}$
refers are called the equivalence class
of local inertial coordinates with the origin at $x$. The neighbourhood
${\cal O}^{\omega}_{x}$ is called a normal neighbourhood.

\section{Local and translocal properties}

One consequence of a local inertial sector of a
physical state would be
the distinction it would draw between the two different ideal sets of
kinematical procedures. Given a physical state $\omega$ and a
point $x\in \cal M$,
consider a local inertial sector $\{\pi^{\omega}|{\cal O}^{\omega}_{x}\}$.
A kinematical procedure in $A\in {\cal A}({\cal O}^{\omega}_{x})$ is called
translocal (or absolute) if it escapes the local action of
the dynamical group $G_{\omega}$ in
${\cal O}^{\omega}_{x}$. In mathematical terms, this is taken to mean
that for an arbitrary
element $\alpha\in G_{\omega}$ we have
$\alpha (A)-A \in {\cal I^{\omega}} ({\cal O}^{\omega}_{x})$. A kinematical
procedure $A\in {\cal A}({\cal O}^{\omega}_{x})$ for which
this condition can not be satisfied is called local\footnote{In reality
there are certain limitations on the applicability of
this definition,
because of the limited accuracy of actual
experiments which makes it impossible to determine the ideal
${\cal I^{\omega}} ({\cal O}^{\omega}_{x})$ exactly. We shall ignore
problems of this type.}.\\
It can be shown that this distinction between local and translocal
kinematical procedures is preserved at the dynamical level of the theory.
In fact we prove the following\\
$Statement$:
For a local (respectively translocal) kinematical procedure
$A\in {\cal A}({\cal O}^{\omega}_{x})$ the corresponding
equivalence class in the sense of (\ref{eq.}) contains local (respectively
translocal)
kinematical procedures only.\\
Consider first the case of a translocal kinematical procedure
$A\in{\cal A}({\cal O}^{\omega}_{x})$.
We show that any kinematical procedure
$B\in{\cal A}({\cal O}^{\omega}_{x})$ which is equivalent to $A$ is
translocal. For an arbitrary element $\alpha$
of the dynamical group $G_{\omega}$ we have $\alpha(A)=A+I$ with
$I\in {\cal I}({\cal O}^{\omega}_{x})$. Since $B\sim A$ we also have
$B=A+I^{\prime}$ with
$I^{\prime}\in {\cal I}({\cal O}^{\omega}_{x})$. It then follows for all
$\alpha\in G_{\omega}$ that
$$
\alpha(B)=\alpha(A)+\alpha(I^{\prime})=A+I+\alpha(I^{\prime})=B-I^{\prime}+
I+\alpha(I).
$$
This together with the invariance of the ideal, relation (\ref{2eq1}), implies
$\alpha(B)-B\in {\cal I^{\omega}} ({\cal O}^{\omega}_{x})$. Thus $B$ is
translocal.
Now consider the case of a local kinematical procedure
$A\in {\cal A}({\cal O}^{\omega}_{x})$. We show that
any kinematical procedure $B\in {\cal A}({\cal O}^{\omega}_{x})$
which is equivalent to $A$ is local.
We have $B=A+I$ with $I\in {\cal I^{\omega}} ({\cal O}^{\omega}_{x})$.
Since $A$ is local there exist an element $\alpha$ of the dynamical group
so that the difference
$\Delta=\alpha(A)-A$ does not lie in ${\cal I^{\omega}} ({\cal O}^{\omega}_{x})$.
It follows that
$$
\alpha(B)=\alpha(A)+\alpha(I)=A+\Delta+\alpha(I)=B-I+\Delta+\alpha(I)
$$
from which one infers that $\alpha(B)-B$ can not be in
${\cal I^{\omega}} ({\cal O}^{\omega}_{x})$.
Thus $B$ is local. \\
From this consideration it follows that the dynamical procedures of a
local inertial sector decompose into two distinct sets, namely the sets
containing all equivalence classes of local
and translocal
kinematical procedures respectively. A member of the first set
(respectively the second set)
is called a local (respectively translocal) dynamical
procedure. \\
We emphasize that this distinction between dynamical procedures
takes the concept of dynamical
activity in a local inertial sector as basic.
A translocal dynamical procedure in a local inertial sector is taken
to be a dynamical
procedure that continually transforms into itself by the local action of
the dynamical group. They correspond to absolute properties of a local
inertial sector.\\
It should be emphasized that the appearance of the translocal
kinematical procedures
in ${\cal A}({\cal O}^{\omega}_{x})$ illustrates a novel effect of the
principle of general covariance. In fact,
any restriction to local kinematical procedures inside a local
inertial sector $\{\pi^{\omega}|{\cal O}^{\omega}_{x}\}$
would be fundamental only to
the extend to which the diffeomorphism group refers only to the
properties inside the normal neighbourhood ${\cal O}^{\omega}_{x}$.
That this is not the case is seen by the following consideration
which furnishes the necessary prerequisite for our subsequent presentations.\\
Consider the identification of points
inside ${\cal O}^{\omega}_{x}$,
made in a system of local inertial coordinates, and
consider a kinematical procedure parametrized
in a local coordinate system outside ${\cal O}^{\omega}_{x}$.
In general a kinematical procedure of this type
characterizes a
nonlocal kinematical property
outside ${\cal O}^{\omega}_{x}$ which does not transform
under the change of the system of local inertial coordinates inside
${\cal O}^{\omega}_{x}$,
so it escapes the local action of the dynamical group in
${\cal O}^{\omega}_{x}$.
Now consider a
diffeomorphism acting entirely outside ${\cal O}^{\omega}_x$.
We shall call a diffeomorphism of this type a gauge
transformation.
The essential point is that it needs only to apply an appropriate
gauge transformation, namely a gauge transformation which has its image
inside ${\cal O}^{\omega}_x$, to convert a nonlocal
kinematical property outside ${\cal O}^{\omega}_{x}$ into a translocal
kinematical procedure inside ${\cal O}^{\omega}_{x}$.
This argument demonstrates that a translocal kinematical procedure
is the image of a non-local kinematical procedure outside
${\cal O}^{\omega}_{x}$ under an appropriate gauge transformation.
Thus, gauge transformations can be applied to generate the totality
of all translocal kinematical procedures inside ${\cal O}^{\omega}_x$ as
the local codifications of the totality of
all nonlocal kinematical procedures outside ${\cal O}^{\omega}_{x}$.
This connection between a local inertial sector and
the associated appearance
of translocal
(absolute) properties
is the distinctly marked conclusion of
the present analysis.\\
At this point a clarifying remark concerning the status of translocal
kinematical procedures with respect to the conventional
quantum field theory appears to be in place. From our presentation one can
immediately observe that, in any theory in which one finds a dynamical
group globally acting on the underlying (space-time) manifold,
there would be no obvious way to introduce (quasi) invariant
kinematical procedures with respect to that group, so
a translocal kinematical procedure would not be obvious in the fundamental
description of the theory. This is specially so in Minkowski-space
quantum field theory with the Lorentz-group playing
the role of a global dynamical group\footnote{The situation would change
if one considers
the embedding of the manifold with a global dynamical group into a larger
manifold without extending the action of the dynamical group. In this case
any kinematical procedure which lies outside the initial manifold
can obviously be interpreted as a (quasi)invariant object with respect
to the action of dynamical group.}.
In particular, in the latter theory the proven statement at the begin of
this section trivializes because
all kinematical procedures becomes essentially local, because they
can not escape the global action of a nontrivial Lorentz-transformation.

\section{The axioms of translocality}

From the scheme presented so far one can immediately infer that
the set of all translocal
dynamical procedures in a local inertial sector
$\{\pi^{\omega}|{\cal O}^{\omega}_{x}\}$ is closed under the algebraic
operations.
This statement
may not have in general an analog with respect to the
local dynamical procedures. Actually, there is a principal
possibility
that a translocal dynamical procedure can be
approximated by finite algebraic
operations on local dynamical procedures.
In such a situation
the dynamical informations
monitored by an actual measurement on a physical system would algebraically
connect both the local and the translocal properties. It is not the objective of
this paper to develop the particular mathematical formalism needed to
describe physics of this sort which is indeed a very
complicated enterprise\footnote{It may be expected that nonunitary
evolution would be the dominating feature of physics of this type.}.\\
The kind of behavior, that we may expect
to occur
for a large class of physical systems
in the generally covariant context,
is that it should categorically not possible to connect the local
and translocal dynamical procedures by a finite (or infinite in an
admissible sense)
algebraic process
in a local inertial sector. Mathematically, this requirement
may be converted into the first axiom of translocality
formulated as the statement:\\
The set of all local dynamical procedures
in a local inertial sector
$\{\pi^{\omega}|{\cal O}^{\omega}_{x}\}$
generates a weakly closed subalgebra of
${\cal A}^{\omega}({\cal O}^{\omega}_{x})$ which has
a trivial intersection\footnote{The difference between a
trivial intersection and an empty intersection is that the former is allowed
to contain multiples of the identity element.} with the algebra of
translocal dynamical procedures inside ${\cal O}^{\omega}_{x}$. \\
This axiom emphasizes the feasibility of a substantial distinction
between the local and translocal properties inside a local inertial
sector.\\
We shall exclusively deal with theories satisfying this axiom. For
such theories the local kinematical procedures in
${\cal A}({\cal O}^{\omega}_{x})$
can be identified with
ordinary local observation procedures (pure description of possible
laboratory measurements) and their corresponding equivalence classes
in ${\cal A}^{\omega}({\cal O}^{\omega}_{x})$
with local observables. The equivalence classes of translocal
kinematical procedures in
${\cal A}^{\omega}({\cal O}^{\omega}_{x})$
correspond to the properties
which do not respond to a local measurement process inside
${\cal O}^{\omega}_{x}$. We
denote the algebra generated by local observables of a normal neighbourhood
${\cal O}^{\omega}_{x}$
by ${\cal A}^{\omega}_{obs}({\cal O}^{\omega}_{x})$. It is considered
as a weakly closed subalgebra of
${\cal A}^{\omega}({\cal O}^{\omega}_{x})$. \\
Particular attention should be directed to the transformation properties
of a local inertial sector $\{\pi^{\omega}|{\cal O}^{\omega}_{x}\}$
under various automorphisms of
${\cal A}^{\omega}({\cal O}^{\omega}_{x})$.
Consider
first the case of an inner-automorphism $\alpha$ of
${\cal A}^{\omega}({\cal O}^{\omega}_{x})$
generated by a translocal dynamical procedure $\cal U$, namely
\begin{equation}
\alpha(A)={\cal U} A {\cal U}^{-1},~~~\forall A\in
{\cal A}({\cal O}^{\omega}_{x}).
\label{4a}\end{equation}
An inner-automorphism of this kind is called a translocal morphism.
The properties of a physical system in the generally covariant context
depends very crucially on the particular way in which a translocal
morphism acts geometrically. The second axiom of translocality assumes
a one to one correspondence between the action of
a translocal morphism and the action of a gauge transformation.
More precisely, this axiom emphasizes that a given translocal
morphisms has a geometric action corresponding to the action of a
gauge transformation and conversely a given gauge transformation has
an algebraic action corresponding to a translocal morphism.\\
Since gauge transformations
are diffeomorphisms acting entirely outside ${\cal O}^{\omega}_{x}$, it follows that
a translocal morphism should not affect the local
observables inside the normal neighbourhood ${\cal O}^{\omega}_{x}$.
This would require
an arbitrary element $\cal U$ of the algebra of the
translocal dynamical procedures to commute with all
local observables of a local inertial sector
$\{\pi^{\omega}|{\cal O}^{\omega}_{x}\}$, namely
\begin{equation}
[A, {\cal U}]=0,~~~\forall A\in
{\cal A}^{\omega}_{obs}({\cal O}^{\omega}_{x}).
\label{4b}\end{equation}
Thus the second axiom of translocality implies that the total activity of
translocal dynamical procedures
inside a local inertial sector can be reduced to the presence
of a (nontrivial) commutant of the algebra of local observables in that
sector. We call it
the translocal commutant of a local inertial sector.
It will be denoted
by $[{\cal A}^{\omega}_{obs}({\cal O}^{\omega}_{x})]^{\prime}$.\\
Using the first axiom of translocality we can establish
a general property of the translocal commutant
$[{\cal A}^{\omega}_{obs}({\cal O}^{\omega}_{x})]^{\prime}$.
We prove, namely, that
$[{\cal A}^{\omega}_{obs}({\cal O}^{\omega}_{x})]^{\prime}$
should have a trivial center: Let us assume the opposite case. Then,
by applying the bicommutant property
$[{\cal A}^{\omega}_{obs}({\cal O}^{\omega}_{x})]^{\prime\prime}=
{\cal A}^{\omega}_{obs}({\cal O}^{\omega}_{x})$, we would get a nontrivial
intersection of the local elements of
${\cal A}^{\omega}_{obs}({\cal O}^{\omega}_{x})$ and the translocal elements
of $[{\cal A}^{\omega}_{obs}({\cal O}^{\omega}_{x})]^{\prime}$.
This is a contradiction to the first axiom of translocality. Thus,
the triviality of the center of
$[{\cal A}^{\omega}_{obs}({\cal O}^{\omega}_{x})]^{\prime}$ becomes
imperative.\\
We may note that
the triviality of the
center of $[{\cal A}^{\omega}_{obs}({\cal O}^{\omega}_{x})]^{\prime}$
may be illustrated as a statement about the
global definiteness of the totality of all (non-local)
complementary properties of a
local inertial sector $\{\pi^{\omega}|{\cal O}^{\omega}_{x}\}$.
In the generally covariant context,
this definiteness seems to be
important in determining the long range
dynamical coupling of a physical state with distant sources.
In particular this global definiteness proves to be
very crucial in forming the algebraic action of
dynamical group $G_{\omega}$ inside a local inertial sector
$\{\pi^{\omega}|{\cal O}^{\omega}_x\}$. To illustrate this point, we note first
that, by assumption, this action
leaves the translocal dynamical procedures in
$\{\pi^{\omega}|{\cal O}^{\omega}_x\}$ unchanged. The most immediate way
to manifestly express this property is to approximate
an element $\alpha\in G_{\omega}$ inside ${\cal O}^{\omega}_x$
by an inner-automorphism of
${\cal A}^{\omega}({\cal O}^{\omega}_{x})$ generated by a corresponding element
$L_{\alpha} \in {\cal A}^{\omega}_{obs}({\cal O}^{\omega}_{x})$, namely
\begin{equation}
\alpha(A)=L_{\alpha} A L_{\alpha}^{-1},~~~
\forall A\in {\cal A}^{\omega}({\cal O}^{\omega}_{x}).
\label{4c}\end{equation}
This relation can be used
to study the nature of the group-operator $L_{\alpha}$. We are
particularly interested in a situation in which
the group-operator $L_{\alpha}$ is uniquely determined by this relation.
In general, this relation leaves us an ambiguity
concerning the choice of the group-operator $L_{\alpha}$. In fact,
with (\ref{4c}) we get the freedom to replace the
group-operator $L_{\alpha}$ by $L_{\alpha} C$, where $C$ is
an arbitrary element in the center
of the translocal commutant
$[{\cal A}^{\omega}_{obs}({\cal O}^{\omega}_{x})]^{\prime}$.
We infer that the triviality
of the center of $[{\cal A}^{\omega}_{obs}({\cal O}^{\omega}_{x})]^{\prime}$,
which was implied by the first axiom of translocality,
appears to be a powerful restriction in order
to characterize the group-operator $L_{\alpha}$.\\
Putting the totality of the translocal dynamical procedures
into the translocal commutant
$[{\cal A}^{\omega}_{obs}({\cal O}^{\omega}_{x})]^{\prime}$
by no means implies that correlations can not occur between
the local observables and the translocal
dynamical procedures inside ${\cal O}^{\omega}_x$.
Indeed, an essential input
is to make an assumption of general nature to characterize the form
of the correlations implied by the activity of the translocal
dynamical procedures. This issue is addressed by formulating the third axiom
of translocality which reflects the impossibility of isolating
the algebra generated by local observables with respect to the dynamical activity of the translocal
commutant.
To arrive at its mathematical
formulation we shall require
that for a physical state $\omega$,
the corresponding vector $\Omega^{\omega}$ in a local inertial sector
$\{\pi^{\omega}|{\cal O}^{\omega}_{x}\}$
be a separating vector for the algebra of local observable
${\cal A}^{\omega}_{obs}({\cal O}^{\omega}_{x})$. This
means that it should not be possible to annihilate the vector
$\Omega^{\omega}$ by elements of ${\cal A}^{\omega}_{obs}({\cal O}^{\omega}_{x})$,
namely
\begin{equation}
A~\Omega^{\omega}=0~\rightarrow A=o,~~~
\forall A\in {\cal A}^{\omega}_{obs}({\cal O}^{\omega}_{x}).
\label{C1}\end{equation}
By the standard theorems of the theory of operator algebras [6]
the above requirement can alternatively be replaced
by the requirement of the cyclicity of the
vector $\Omega^{\omega}$ with respect to the translocal commutant
$[{\cal A}^{\omega}_{obs}({\cal O}^{\omega}_{x})]^{\prime}$.
In this formulation the third axiom of translocality
emphasizes the distinguishing role played by translocal
dynamical procedures inside a local inertial sector in singling
out a dense subset of the corresponding Hilbert space.

\section{Commutant duality}

In reality, the more informations which should, in principle, be available
in the form of correlations between local observables and translocal
dynamical procedures
has a significant effect on the effective description
of the short-distance behavior of the underlying theory. To understand
this effect, one has to extrapolate the physical informations carried
by the members of the translocal commutant to the short-distance characteristics
of a local inertial sector. This issue can be addressed by formulating
a duality principle which, in essence, connects the long distance properties
of states with their corresponding short-distance counterparts using
a gauge transformation:\\
Given a local inertial sector
$\{\pi^{\omega}|{\cal O}^{\omega}_{x}\}$, we call any neighbourhood
${\cal O}_{x}\subset {\cal O}^{\omega}_{x}$ of the point $x$ which is invariant under the geometric action
of the dynamical group $G_{\omega}$ an invariant neighbourhood of the
origin.\\
We argue that to any local inertial sector
$\{\pi^{\omega}|{\cal O}^{\omega}_{x}\}$
one can
assign a characteristic invariant
neighbourhood of the origin. By definition, the origin $x$ is invariant
under the geometric action of the dynamical group.
Thus, one needs only to pass from the origin to one of its neighborhoods
${\cal O}_x$
on which the action of the dynamical group remains still arbitrarily close to
the identity such that no local observable can properly
be affiliated to ${\cal O}_x$.
The operational way to
achieve this is as follows: One may start with a contracting sequence
of neighborhoods ${\cal O}_{x}^{\lambda}\subset {\cal O}^{\omega}_{x}$
of the point $x$
\begin{equation}
{\cal O}_{x}^{\lambda+1}\subset {\cal O}_{x}^{\lambda},
\label{co1}\end{equation}
which is ideally taken to shrink to the
point $x$ as $\lambda\rightarrow \infty$, and continue to truncate
the sequence at some sufficiently large $\lambda$. The prefix
'sufficiently large' characterizes an index $\lambda$
for which the set of numbers
$$
|\omega^{\prime}(\alpha(A))-\omega^{\prime}(A)|,~~~
A\in {\cal A}^{\omega}({\cal O}_{x}^{\lambda})
$$
taken for all states $\omega^{\prime}$ in the folium
${\cal F}^{\omega}$ of $\omega$ and for all elements $\alpha$ of the
dynamical group $G_{\omega}$, remains
smaller than a characteristic nonvanishing small number $\epsilon$
characterizing the limited accuracy of the local measurements.
In this way
the correspondence between a local inertial sector and a characteristic
invariant
neighbourhood of the origin may be established. For this neighbourhood
we use the name the continuous image of the origin.\\
Our objective is now to apply the second axiom of translocality to derive
a duality principle which emphasizes the significance of
translocal dynamical procedures for modelling the algebra
corresponding to the continuous image of the origin in a local inertial
sector.
Consider a gauge-transformation  $\sigma$
in a local inertial sector, which according to the second
axiom of translocality, has the algebraic action
corresponding to a translocal morphism.
Given a local inertial sector $\{\pi^{\omega}|{\cal O}^{\omega}_{x}\}$ it is
always possible to obtain a representation in the equivalence class of
$\pi^{\omega}$ by applying a gauge transformation $\sigma$
to ${\cal A}^{\omega}({\cal O}^{\omega}_{x})$.
Let us now consider
a gauge transformation $\sigma$ which sends the totality
of points outside
${\cal O}^{\omega}_{x}$ into the continuous image of the origin
inside a local inertial sector $\{\pi^{\omega}|{\cal O}^{\omega}_{x}\}$.
It follows that for the image of the translocal commutant under $\sigma$
we can establish the inclusion
property
\begin{equation}
\sigma [{\cal A}_{obs}^{\omega}({\cal O}^{\omega}_{x})]^{\prime}\subset
{\cal A}^{\omega}({\cal O}_{x})
\label{co3} \end{equation}
which holds for any neighbourhood
${\cal O}_{x}\subset {\cal O}^{\omega}_{x}$, which contains
the continuous image of the origin as a proper subset. The inclusion property
(\ref{co3}) tells us that the gauge-transformation $\sigma$ can be used
to affiliate the translocal commutant into the continuous image of
the origin. Sine gauge transformations are symmetry operations
inside a local inertial sector we infer that
by restricting  a state to the continuous image of the origin the folium
of $\omega$ becomes indistinguishable from the set of normal
states over the translocal
commutant. This is the expression of what we call the commutant duality.\\
We should emphasize that the geometric gauge transformations
underlying the formulation
of commutant duality is a novel feature of the principle of general covariance
and can not be exemplified in conventional models of quantum field
theory with no geometric gauge group.
Since the geometric gauge transformations can, in principle, be used to affiliate
the translocal commutant to any open region inside the normal
neighbourhood ${\cal O}^{\omega}_{x}$, one can generally say that the theory
deals profoundly with two different phases inside a local inertial
sector, depending on whether the local or the translocal properties
are considered as primary properties. Once this has been recognized, then
the investigation of a possible symmetry between these two distinct
phases appears to be a problem of direct physical relevance.
This symmetry, which can generally be termed
under the name of `duality', needs the study of those coordinate
transformations exchanging the local and translocal dynamical procedures
which are related, in a specific model, to different sets of dynamical variables.
One can generally expect that the formulation
of this symmetry would reflect new geometric
gauge invariance which is not visible inside a
local inertial sector. It is needless to say that
such a development would also shed new light on the symmetry behind the
currently discussed duality of supersymmetric gauge
theory\footnote{See [7] and
references therein.} and string theory.\\
Our last remark in this section concerns the notion of quantum equivalence
principle. There exists a formulation of this principle in the framework
of quantum field theory in curved space which takes the correspondence
between the leading short-distance singularity of states and the corresponding
singularity of the vacuum in Minkowski space [8][2] as basic.
In the present context,
the commutant duality requires a profoundly smooth
short-distance behavior, so there is the need to reformulate the quantum
equivalence principle in a different way. This formulation
is implied by the commutant duality itself. In fact, combining it
with the third axiom of translocality
it follows that
the state-vector  $\Omega^{\omega}$ can be considered as a cyclic vector
for any algebra
${\cal A}^{\omega}({\cal O}_{x})\subset
{\cal A}^{\omega}({\cal O}^{\omega}_{x})$,
for which the neighbourhood ${\cal O}_{x}$
contains the continuous image of the origin as a proper subset.
This cyclicity property establishes an exact correspondence between
the structure of correlations of the state $\omega$ in a local inertial sector
and that of the vacuum state in Minkowski-space\footnote{Actually,
in Minkowski-space there is
a general result, obtained by
Reeh and Schlieder, which states that the vacuum
is cyclic not only for the whole algebra but also for the algebra
of any open region [1]}. We may, therefore, take this correspondence,
which is implied by the commutant
duality, as a coded form of a quantum equivalence principle.

\section{Classical properties}

We analyze now the consequence of commutant duality
in an idealized limit which destroys the algebraic informations
of the translocal commutant in a local inertial sector.
At this level of description
a state is unable to monitor the exact form of
all conceivable correlations between the local observables and the
individual members
of the translocal commutant and the description of a state is transferred
to a positive linear functional over the algebra of local observables in
a local inertial sector. This
corresponds to the conventional description of states in quantum field
theory.
However, the essential point is that, at such a level
of description, the ignorance concerning
the accurate form of algebraic informations contained
in the translocal commutant implies a structural dependence of
the short-distance behavior of the underlying theory on classical properties.
We have to clarify this.\\
Given a local inertial sector $\{\pi^{\omega}|{\cal O}^{\omega}_{x}\}$,
we may ideally transfer          `
the description of the state $\omega$ to a positive linear functional
over the algebra of local observables in that sector. The question
we shall address is how this change of the level of description will
alter the nature of the translocal commutant. The resolution is
quite immediate.
Indeed,
the inclusion relation (\ref{co3})
implies that
the translocal commutant can then be approximated by
a commutative algebra lying in the
center of any subalgebra
${\cal A}^{\omega}_{obs}({\cal O}_{x})\subset
{\cal A}^{\omega}_{obs}({\cal O}^{\omega}_{x})$
for which the neighbourhood
${\cal O}_{x}\subset {\cal O}^{\omega}_{x}$ contains
the continuous image of the origin as a proper subset.
In this way the emergence of classical
properties in a local inertial sector may be an irreducible feature
of the theory if we transfer to the conventional
level of the description of a state in quantum field theory.

\section{Concluding remarks}

In this paper we have discussed the impact of the principle of
general covariance on the algebraic framework of quantum field
theory. At first sight the implementation of this principle seems to create
confusion concerning a substantial
identification of local properties. We have
proposed a tentative resolution of the problem which takes
the dynamical activity in a local inertial sector as basic.
However, the principle of general covariance implies that the
set of all local properties in a local inertial sector may not be considered
as a completed totality. The notion of translocality was introduced to
address this issue.
In our approach there is an effective crossover from
local properties to the translocal properties, once the short-distance scaling
is performed inside a local inertial sector.
This interrelation of short-distance scaling
with the translocal properties which
is implied by the commutant duality may be of particular
importance for expressing Mach's principle [9] within the framework of
quantum field theory [5].
In particular it emphasizes that the short-distance property of
quantum field theory in the generally covariant context is profoundly
different from ordinary quantum field theory. Remarkably,
this is especially so for
an important class of currently discussed theories generally termed by string theory.
We can not at the present understand how an exemplification
of the general principles of generally covariant quantum field theory
in a model can be related to string theory. But, nevertheless it
can be expected that for the unification of
quantum field theory with certain features of string theory the
commutant
duality may have a vital role to play.\\
The next point implied by commutant duality concerns the transition
to the conventional description of states in quantum field theory.
On this level of description the dominant structure of a generally
covariant quantum field theory has been
recognized to be the occurrence of classical
properties. It is an interesting subject to analyze
the interrelation of such classical properties
with the classical space-time metric of general
relativity. We hope to address the issue elsewhere.\\\\
{\bf Acknowledgment}                \\
The Author would like to specially acknowledge the financial support of
the Office of Scientific Research of Shahid Beheshti University.
Thanks are also due to an anonymous referee for useful suggestions.\\\\

{\bf References}\\\\
\begin{tabular}{r p{18 cm}}
1.& Haag R, Local Quantum Physics, Springer (1992) \\
2.& Fredenhagen K and Haag R, Commun. Math. Phys. 108, 91 (1987)\\
3.& Salehi H, Class. Quantum Grav. 9 2557-2571 (1992)\\
4.& Salehi H, Inter.J.Theo.Phy. 36, 9, (1997)\\
5.& Salehi H, Inter.J.Theo.Phy. 36, No.4, (1998)\\
6.& Bratteli O, Robinson D, Operator Algebras and Quantum Statistical
Mechanics I, \\
~ & Springer (1979)\\
7.& Seiberg N, (The Power of duality, Exact results in 4d susy field theory)
hep-th/9506077\\
8.& Haag R, Narnhofer H and Stein U, Commun. Math. phys. 94, 219, (1984)\\
9.& Brans C, Dicke R H, Phys. Rev.124, 925-935 (1961)\\

\end{tabular}

\end{document}